\documentclass[twocolumn, pra, aps, superscriptaddress, longbibliography, showpacs, amsmath, amssymb, floatfix]{revtex4-1}                                                                                         
\usepackage{graphicx}
\usepackage{amssymb}
\usepackage{amsmath}
\usepackage{epsfig}
\usepackage{color}
\usepackage{mathtools}
\usepackage[colorlinks,linkcolor=blue,anchorcolor=blue,citecolor=blue,urlcolor=blue]{hyperref}
\usepackage{physics}
\usepackage{bm}
\begin{document}

\title{Exact dimer phase with anisotropic interaction for one dimensional magnets}
\author{Hong-Ze Xu}
\affiliation{CAS Key Laboratory of Quantum Information, University of Science and Technology of China, Hefei, 230026, China}
\author{Shun-Yao Zhang}
\affiliation{CAS Key Laboratory of Quantum Information, University of Science and Technology of China, Hefei, 230026, China}
\author{Guang-Can Guo}
\affiliation{CAS Key Laboratory of Quantum Information, University of Science and Technology of China, Hefei, 230026, China}
\affiliation{Synergetic Innovation Center of Quantum Information and Quantum Physics, University of Science and Technology of China, Hefei, Anhui 230026, China}
\affiliation{CAS Center For Excellence in Quantum Information and Quantum Physics,  University of Science and Technology of China, Hefei, Anhui 230026, China}
\author{Ming Gong}
\email{gongm@ustc.edu.cn}
\affiliation{CAS Key Laboratory of Quantum Information, University of Science and Technology of China, Hefei, 230026, China}
\affiliation{Synergetic Innovation Center of Quantum Information and Quantum Physics, University of Science and Technology of China, Hefei, Anhui 230026, China}
\affiliation{CAS Center For Excellence in Quantum Information and Quantum Physics,  University of Science and Technology of China, Hefei, Anhui 230026, China}
\date{\today }

\begin{abstract}
We report the exact dimer phase, in which the ground states are described by product of singlet dimer, in the extended XYZ model by 
generalizing the isotropic Majumdar-Ghosh model to the fully anisotropic region. We demonstrate that this phase can be realized even in models when antiferromagnetic interaction along one of 
the three directions. This model also supports three different ferromagnetic (FM) phases, denoted as $x$-FM, $y$-FM and $z$-FM,
polarized along the three directions. The boundaries between the exact dimer phase and FM phases are infinite-fold degenerate. The 
breaking of this infinite-fold degeneracy by either translational symmetry breaking or $\mathbb{Z}_2$ symmetry breaking leads to exact dimer phase and FM phases, respectively. Moreover, the boundaries between the three FM phases are critical with central charge 
$c=1$ for free fermions. We characterize the properties of these boundaries using entanglement entropy, excitation gap, and 
long-range spin-spin correlation functions. These results are relevant to a large number of one dimensional magnets, in which 
anisotropy is necessary to isolate a single chain out from the bulk material. We discuss the possible experimental signatures 
in realistic materials with magnetic field along different directions and show that the anisotropy may resolve the disagreement 
between theory and experiments based on isotropic spin-spin interactions. 
\end{abstract}
\maketitle

\section{Introduction}
\label{sec-introduction}

The spin models for magnetism are basic topics in modern solid-state physics and condensed matter physics 
\cite{Auerbach2012interacting}. In these models, only a few of them mostly focused on low dimensions, 
can be solved exactly. In general, we may categorize these solvable models into two different groups according to 
the methods these models are solved. 

In the first group, the models can be solved exactly using some mathematical techniques based on their symmetries 
\cite{baxter2016exactly} and the dual relation between fermions and spins. Typical examples are the transverse Ising model, 
the XY model, the XXZ model \cite{amico2008entanglement, pfeuty1970one, cheng2018symmetry}, 
the XYZ model \cite{Cao2014spinXYZ,ercolessi2011essential,ercolessi2013modular}, and the toric code 
model \cite{kitaev2003fault, kitaev2006anyons}. Here, the XY model and Ising model can be mapped to the non-interacting 
$p$-wave superconducting model by a non-local Jordan-Wigner transformation, which can then be solved by a unitary transformation 
in the momentum space \cite{batista2001generalized,tzeng2016entanglement,karbach2005spin,yao2011robust,mukherjee2007quenching}. 
The XXZ model is a prototype model for the exact calculation by the Bethe-ansatz approach. In combination 
with the Jordan-Wigner transformation, the XXZ model is mapped to the interacting Hubbard model, for which reason some of the 
Hubbard models may also be solved using the Bethe-ansatz approach by Lieb and Wu \cite{Lieb1968Absence}. 
The XYZ model can also be solved analytically by the off-diagonal Bethe-ansatz method \cite{Cao2014spinXYZ} and modular transformations method \cite{ercolessi2011essential,ercolessi2013modular}. The Bethe-ansatz approach has broad applications in many-body physics.
With the above approaches, their spectra, partition function
and correlation functions of these models can be obtained exactly. Recently, the spinon excitations in these models have 
been directly measured in experiments by neutron diffraction \cite{mourigal2013fractional,
wang2018experimental,wang2015spinon}. In the two dimensional models, the Kitaev toric code model can be solved exactly by 
considering the gauge symmetries in each plateau \cite{kitaev2003fault, kitaev2006anyons}. These solvable models have also 
played an essential role in the understanding of the non-equilibrium dynamics, phase transitions, and entanglement in the 
many-body systems \cite{mukherjee2007quenching, osterloh2002scaling, vidal2003entanglement,bonnes2014light,bertini2016transport}. 

\begin{figure}
\centering
\includegraphics[width=0.50\textwidth]{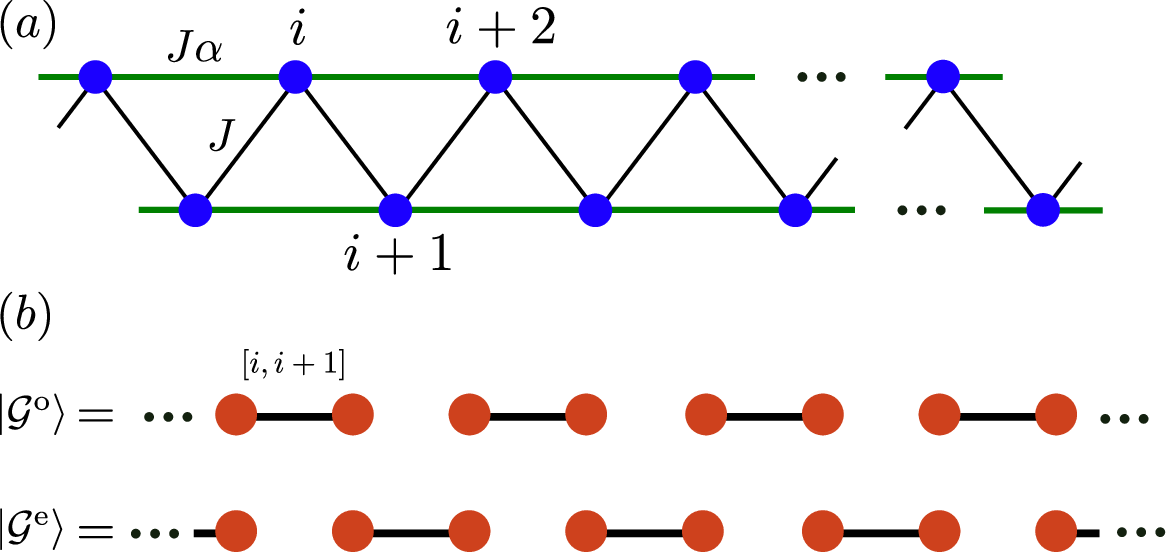}
	\caption{(a) The model in Eq. \ref{eq-H} with nearest $J$ and next-nearest-neighbor interaction $J\alpha$. 
	(b) The schematic illustration of the two exact dimer states, in which each singlet dimer is represented by a 
	solid bound (see the exact definition of the wave function in Eq. \ref{eq-wf}). }
\label{fig-fig1}
\end{figure}

In the second group, which is most relevant to the research in this work, only the ground states (GSs) of the Hamiltonian can be obtained. For example, in the most representative spin-1/2 
Majumdar-Ghosh (MG) model \cite{majumdar1969next,majumdar1969next2,chhajlany2007entanglement, liu2011quantum,lavarelo2013localization,ramkarthik2013entanglement}, which reads as 
\begin{equation}
	H_{\text{MG}} =  J \sum_i^L (h_{i, i+1}^0 + \alpha h_{i, i+2}^0),
	\label{eq-MG}
\end{equation}
with
\begin{equation}
	h_{ij}^0 = s_i^x s_j^x + s_i^y s_j^y + s_i^z s_j^z = {\bf s}_i \cdot {\bf s}_j, \quad \alpha = {1 \over 2}.
\end{equation}
This model can be obtained from the Fermi-Hubbard by second-order exchange interaction, thus $J > 0$ for anti-ferromagnetic 
interaction. The GSs of the above model can be expressed exactly as the product of singlet dimers. This model preserves 
the three $\mathbb{Z}_2$ symmetries by defining $s_i^x \rightarrow -s_i^x$, $s_i^y \rightarrow -s_i^y$ and $s_i^z \rightarrow 
s_i^z$ and its index rotation. Using the above Jordan-Wigner transformation, the next-nearest-neighbor interaction
and the coupling along the $z$-direction can yield complicated many-body interaction, thus this model can not be solved 
analytically using the approach in the first group. However, the GSs can be constructed using some special tricks 
with the help of the projector operators. Let us define $\boldsymbol{\Pi} = {\bf s}_i + {\bf s}_{i+1} + {\bf s}_{i+2}$, 
with ${\bf s}^2_i = (s_i^x)^2 + (s_i^y)^2 + (s_i^z)^2 = \frac{3}{4}$, we can obtain
\begin{equation}
\boldsymbol{\Pi}^2 = \frac{9}{4} + 2({\bf s}_i \cdot {\bf s}_{i+1} + {\bf s}_i \cdot {\bf s}_{i+2} +{\bf s}_{i+1} \cdot {\bf s}_{i+2}) = S(S+1),
\end{equation}
with $S=\frac{1}{2}$ or $\frac{3}{2}$ from the decoupling ${1 \over 2} \otimes {1 \over 2} \otimes {1 \over 2} = {1\over 2} \oplus {1\over 2} \oplus {3 \over 2}$. The above
result means that the total spin space can be decoupled into three different irreducible representations. Let us define the corresponding projectors for these 
subspaces as $P_S(i, i+1, i+2)$, then we have
\begin{equation}
	{\bf s}_i \cdot {\bf s}_{i+1} + {\bf s}_i \cdot {\bf s}_{i+2} +{\bf s}_{i+1} \cdot {\bf s}_{i+2} = \frac{3}{2} P_{\frac{3}{2}}(i, i+1, i+2) - \frac{3}{4}.
\end{equation}
The projectors have the feature that $P_S(i, i+1, i+2) P_{S'}(i, i+1, i+2) = \delta_{SS'} P_S(i, i+1, i+2)$ and $\langle \psi| P_S(i, i+1, i+2)|\psi\rangle \ge 0$ for any wave function. Then MG model of Eq. \ref{eq-MG} can be rewritten as
\begin{equation}
\begin{aligned}
H_{\text{MG}} & = J\sum_i^L \frac{3}{4} [P_{\frac{3}{2}}(i,i+1,i+2) - \frac{1}{2}] \\
	& = -\frac{3}{8}J L + \frac{3J}{4} \sum_{i} P_{\frac{3}{2}}(i,i+1,i+2).
	\label{eq-Ps}
\end{aligned}
\end{equation}
Here the project in the singlet subspace $P_{1/2}$ is absent from the Hamiltonian. The GSs energy of $H_{\text{MG}}$ is 
given by $-\frac{3}{8}L$, which means that for any $i$, the ground state $|\mathcal{G}\rangle$ should satisfy
\begin{equation}
P_{\frac{3}{2}}(i,i+1,i+2)|\mathcal{G}\rangle = 0.
\end{equation} 
This constraint requires $J > 0$; otherwise, the triplet state(s) should have much lower energy. To this condition, there 
must be a singlet in the three adjacent sites for the eigenvectors of $P_{1/2}(i, i+1, i+2)$. Mathematically, the two exact 
dimer GSs can be written as
\begin{equation}
    |\mathcal{G}^{\text{e}}\rangle = \prod_{2n} [2n, 2n+1], \quad 
    |\mathcal{G}^{\text{o}}\rangle = \prod_{2n} [2n-1, 2n],
    \label{eq-wf}
\end{equation}
where $[i,i+1] = {1\over \sqrt{2}}|\uparrow_i \downarrow_{i+1} - \downarrow_i \uparrow_{i+1}\rangle$ represents the singlet 
dimer between neighboring sites (see Fig. \ref{fig-fig1} (b) with solid bounds). This idea was generalized to the 
Affleck-Kennedy-Lieb-Tasaki (AKLT) model in a spin-1 chain with
\begin{equation}
	H^{\text{AKLT}} = J \sum_i {\bf s}_i \cdot {\bf s}_{i+1} + {1 \over 3} ({\bf s}_i \cdot {\bf s}_{i+1})^2,
\end{equation}
which was one of the most important models for the Haldane phase \cite{Affleck2004,santos2011negativity,morimoto2014z,wei2011affleck}. The degeneracy of the GSs of this model can 
be solved using the above constructive approach. The AKLT model is also one of the basic models for the searching of symmetry protected topological (SPT) phases, which are frequently searched by the above construction method.

The MG model may be relevant to a large number of one dimensional magnets in experiments in solid materials, such as CuGeO$_3$ \cite{hase1993observation,o2017vibronic,meibohm2018comparison}, TiOCl \cite{rotundu2018enhancement,zhang2014transformation}, Cu$_3$(MoO$_4$)(OH)$_4$ \cite{lebernegg2017frustrated}, 
DF$_5$PNN \cite{inagaki2017phase}, (TMTTF)$_2$PF$_6$ \cite{pouget2017inelastic}, (o-Me$_2$TTF)$_2$NO$_3$ \cite{jeannin2018decoupling}
and MEM(TCNQ)$_2$ \cite{poirier2013charge}, {\it etc}. In these materials, the lattice constant along one of the directions is much smaller than the other two directions, rending the couplings between the 
magnetic atoms along the shortest lattice constant direction is much stronger than along the other two directions,
 giving rise to one dimensional magnets. To date, most of these candidates are explained based on the isotropic spin 
 models. It was found that these isotropic models are insufficient to understand all results in experiments 
 \cite{miljak2005anisotropic, wang2013single, Kremer1995Anisotropic}. 

There are two major starting points for this work. Firstly, we hope to generalize the physics discussed in the isotropic models 
to the fully anisotropic models, which may contain some beautiful mathematical structures. 
Secondly, we hope to provide a possible model to study the one dimensional magnets observed in experiments, as 
above mentioned, which contain some more possible tunable parameters while the fundamental physics is unchanged. 
In other words, the physics based on isotropic interaction can be found in some more general Hamiltonians. 
Our model harbors not only the exact dimer phase, but also three gapped ferromagnetic (FM) 
phases, denoted as $x$-FM, $y$-FM and $z$-FM, according to their magnetic polarization directions. We can determine their 
phase boundaries analytically based on a simplified model assuming exact dimerization. We find that the boundaries between exact dimer phase and FM phases are infinite-fold degenerate, while the boundaries between the FM phases are gapless and critical 
with central charge $c=1$ for free fermions. Thus these two phases represent either the translational symmetry breaking or 
the $\mathbb{Z}_2$ symmetry breaking from the infinite-fold degenerate boundaries. We finally discuss the relevance of our 
results to one dimensional magnets and present evidences to distinguish them in experiments, showing that it explains 
both the exact dimer phase and the anisotropic susceptibility, which are simultaneously obtained in experiments. 

This manuscript is organized as the following. In section \ref{sec-model}, we present our model for the generalized MG model with anisotropic XYZ interaction. In section \ref{sec-exactdimer}, we present a method to obtain the exact dimer phase and the associated 
phase boundaries. We will map out the whole phase diagram based on this analysis and confirm our results with high accuracy using
exact diagonalization method and density matrix renormalization group (DMRG) method. In section \ref{sec-fm}, we will discuss the 
three ferromagnetic phases. In sections \ref{sec-model} to \ref{sec-fm}, 
we mainly discuss the physics in the MG point with $\alpha = 1/2$ 
for the sake of exact solvability; however, the similar physics will be survived even away from this point. In 
section \ref{sec-app}, we will show how this model can find potential applications in some of the one dimensional magnets 
away from the MG point. Finally, we conclude in section \ref{sec-conc}. Details about the phase boundaries and the general 
theorem will be presented in the Appendix.

\begin{figure}
\centering
\includegraphics[width=0.50\textwidth]{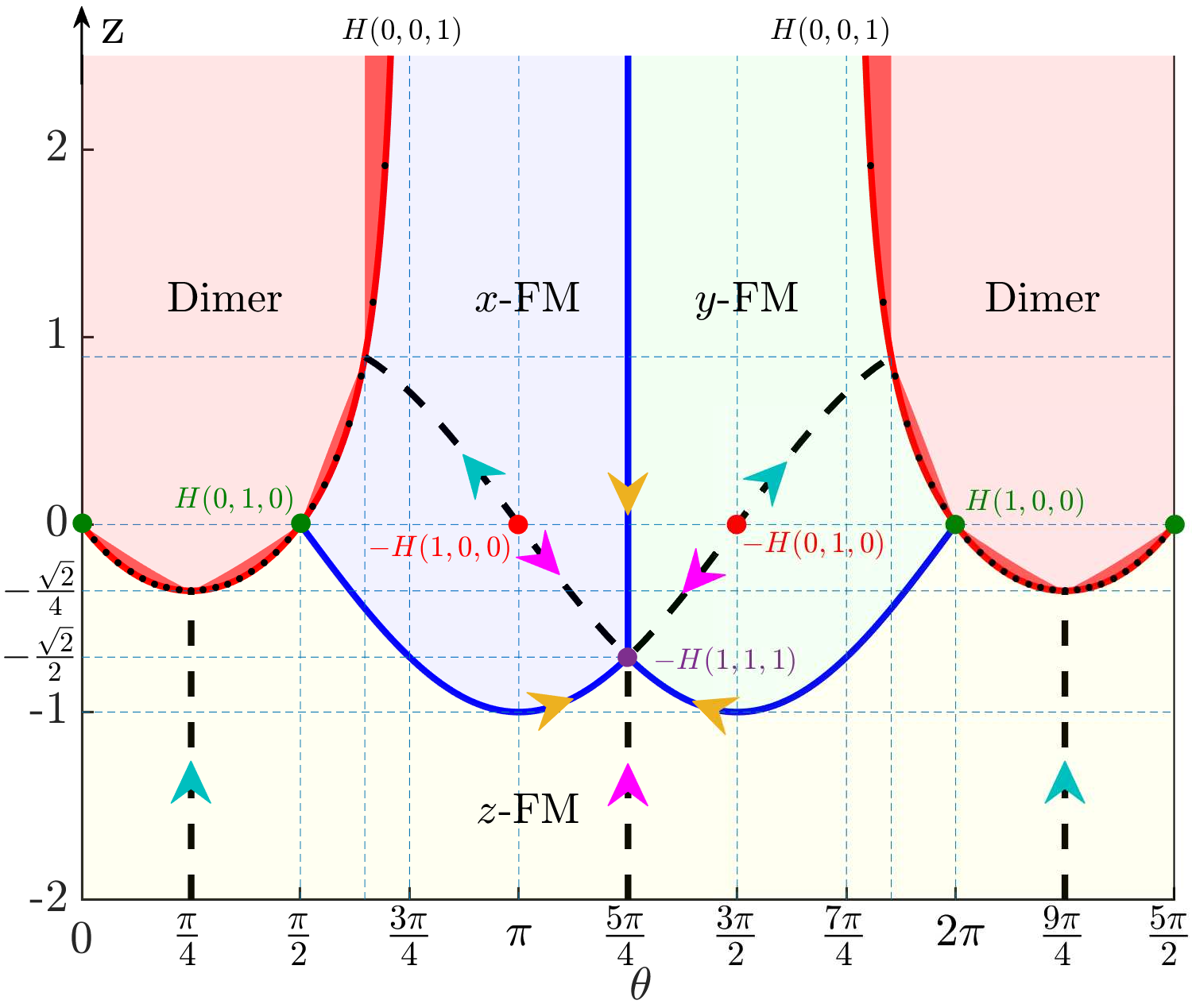}
	\caption{Phase diagram for the fully anisotropic XYZ model in Eq. \ref{eq-H}. We have assumed $x = \cos(\theta)$ and $y = \sin(\theta)$. The phase boundaries between exact dimer phase and FM 
	phases are determined by Eq. \ref{eq-b1}, while the dots are boundaries determined by order parameters, with absolute 
	difference $|z_c - z_\text{ex}|$ ($z_\text{ex}$ is the exact boundary given by Eq. \ref{eq-b1}) less than 
	$3.0 \times 10^{-4}$. In the exact dimer phase, the deep red regions can not be explained by mixing of two anisotropic 
	dimer models; see Eq. \ref{eq-mixed}. The classical limits are denoted as $H(1, 0, 0)$, $H(0,1,0)$ and 
	$H(0,0,1)$ and the dashed lines are conditions for exact FM states.}
\label{fig-fig2}
\end{figure}

\section{Model and Hamiltonian }
\label{sec-model}

We consider the following spin-1/2 model directly generalized from the isotropic MG model (see Fig. \ref{fig-fig1} (a)), 
\begin{equation}
    H(x, y, z) = J\sum_i^L h_{i, i+1} + \alpha h_{i, i+2},
	\label{eq-H}
\end{equation}
where $\alpha = {1 \over 2}$ (MG point) and $J > 0$. For the anisotropic Heisenberg interaction, we have 
\begin{equation}
h_{i,j} = x s_i^x s_{j}^x + y s_i^y s_{j}^y  + z s_i^z s_{j}^z,
\end{equation}
with $x,y,z \in \mathbb{R}$. Hereafter, we let $J = 1$, unless specified. 
The case when $x=y=z > 0$ corresponds to the well-known MG model with exact dimer phase based on isotropic interaction  \cite{majumdar1969next,majumdar1969next2}. 
Anisotropy can be introduced to this model by letting $x = y > 0$,  in which when $z > -x/2$ the GSs are also exactly dimerized 
with XXZ interaction \cite{kanter1989exact,nomura1993phase,gerhardt1998metamagnetism}. 

There are several ways to extend this model to more intriguing and more realistic conditions, considering the possible anisotropy 
in real materials. For example, in the presence of some proper long-range interactions \cite{kumar2002quantum}, the GSs can still 
be exactly dimerized using the constructive approach in section \ref{sec-introduction}. When this model is generalized to 
integer spins, it may support SPT phases \cite{hindmarsh2016new,politis2016non,gils2009collective,furukawa2012ground}. However, in 
the presence of anisotropy as discussed above, which can not be solved analytically, the physics is largely unclear. 

\section{Exact dimer phase}
\label{sec-exactdimer}

Our phase diagram of the exact dimer phase for Eq. \ref{eq-H} is presented in Fig. \ref{fig-fig2}. This phase has the 
advantage to be determined exactly with even small lattice sites with periodic boundary condition (PBC). We will confirm 
the analytical phase boundary with high accuracy using numerical methods. 

\subsection{Phase boundary}

\begin{figure}
\centering
\includegraphics[width=0.50\textwidth]{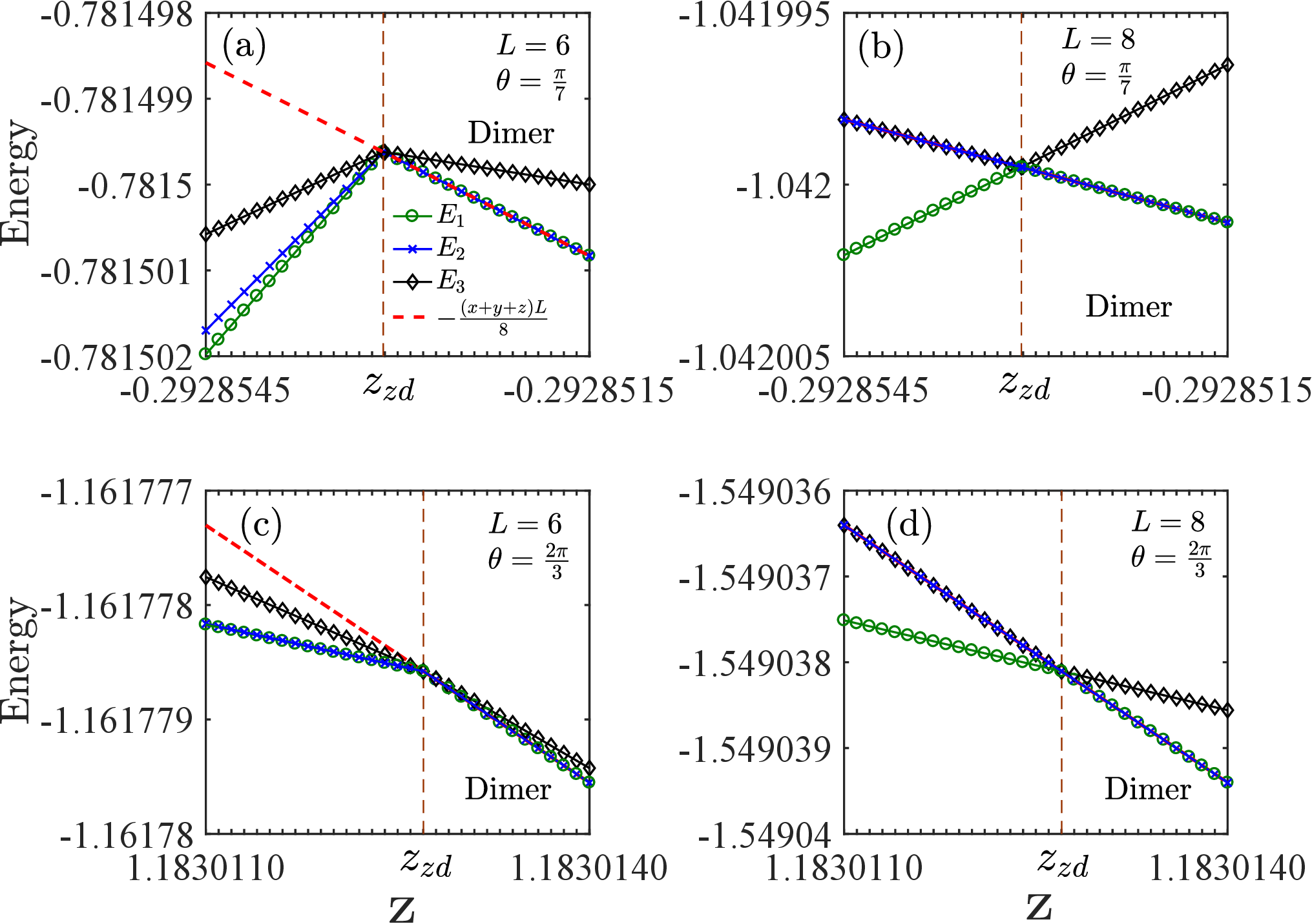}
	\caption{Energy spectra of the lowest three levels for small lattice sites with PBC based on exact diagonalization method. 
	(a) - (b) show the exact dimer phase boundary ($z_{\text{ex}}=-xy/(x+y) = -0.2928531$) at $\theta=\frac{\pi}{7}$ with $L=6$ and $L=8$. (c) - (d) show the exact dimer phase boundary ($z_{\text{ex}}= -xy/(x+y)=1.1830127$) at $\theta=\frac{2\pi}{3}$ with $L=6$ and $L=8$. In the exact dimer phase, the GS energy of the two-fold degenerate states is given by Eq. \ref{eq-GS_dimer} (red dashed lines). }
\label{fig-fig3}
\end{figure}

The exact dimer states in Eq. \ref{eq-wf} are independent of system parameters, indicating 
that it is also exact even in a finite system. To this end, we consider the simplest case with 
$L = 4$ with Hamiltonian as 
\begin{equation}
H_4 = h_{12} + h_{23} + h_{34}  + h_{41} + \alpha[h_{13} + h_{24} + h_{31} + h_{42}].
\end{equation}
This model can be solved analytically with eigenvalues given below
\begin{equation}
\left\{
\begin{aligned}
    &E_{1-3} = -\frac{x}{2}, \quad E_{4-6} = -\frac{y}{2}, \quad E_{7-9} = -\frac{z}{2},   \\
    &E_{10} = \frac{3x}{2}, \quad E_{11} = \frac{3y}{2}, \quad E_{12} = \frac{3z}{2},   \\
	&E_{13-14} =\frac{1}{2}(x+y+z \pm \sqrt{x^2+y^2+z^2 -xy - yz - zx}),   \\
    &E_{15-16}^\text{dimer} = -\frac{x+y+z}{2}.
\end{aligned}
\right.
    \label{eq16}
\end{equation}
The last two states with two-fold degeneracy correspond to the exact dimer phase with eigenvectors in the form of 
Eq. \ref{eq-wf}. One can verify that this model can be solved analytically only at the MG point with $\alpha = 
1/2$. To request the exact dimer states have the 
lowest energy, we request $E_{15-16}^{\text{dimer}} < E_i$ for $i = 1 - 14$, which yields
\begin{equation}
    x+y+z > 0,  \quad xy +yz + zx > 0.
\label{eq-b1}
\end{equation}
This is the major phase boundary determined for the exact dimer phase (see boundaries in Fig. \ref{fig-fig2}). Let's assume $x+y > 0$, then the second equation yields the exact phase boundary 
\begin{equation}
	z > z_\text{ex} = -\frac{xy}{x+y}.
\label{eq-b2}
\end{equation}
The same boundary can be obtained for $L = 6$ and $8$ with high accuracy from the eigenvalues and ground state degeneracy 
(see Fig. \ref{fig-fig3}). By this result, the GSs energy for the exact dimer phase for a chain with length $L$ ($L$ is an even 
number), following Eq. \ref{eq16}, is given by
\begin{equation}
    E_g^\text{dimer} = -\frac{(x+y+z)L}{8}.
    \label{eq-GS_dimer}
\end{equation}
This result naturally includes the previously known results in the MG model with $x=y=z > 0$ \cite{kumar2002quantum} and the 
extended XXZ model with $x=y > 0$ and $z > -x/2$ \cite{kanter1989exact,nomura1993phase,gerhardt1998metamagnetism}. The 
accuracy of this boundary will be checked by the order parameters in the next subsection. 

As discussed in the section of the introduction, the ferromagnetic interaction with $J > 0$ is essential for
the exact dimer states; otherwise, the triplet state is more energetically favorable (see Eq. \ref{eq-Ps}).
Here, Eq. \ref{eq-b1} can lead to an interesting conclusion beyond this criterion that the exact dimer states can 
be found in the anisotropic model with some kind of antiferromagnetic interaction. For the three parameters in Eq. \ref{eq-H}, we find that this exact dimer phase can be realized when only one of the anisotropic parameters 
is negative valued. It can be proven as follows. 
Let $y$ and $z$ be negative values, then $x > |y| + |z| > 0$. 
However, the second condition in Eq. \ref{eq-b1} means ${1\over x} > {1 \over |y|} + {1\over |z|}$. The multiply of these two 
inequalities yields an obvious contradiction. For the case with two negative parameters, one may compute $-H$, which may support exact dimer states in its GSs. In this way, the highest levels of $H$ can be exactly dimerized
when Eq. \ref{eq-b1} is satisfied. 

Then, how to understand the phase boundary in Eq. \ref{eq-b1}? Whether this boundary contains some nontrivial region
that can not be explained by the known results in the previous literature? To this end, we first need to prove another 
model for the exact dimer phase. For $z = 0$ and $x > 0$, $y>0$, let us define 
\begin{equation}
H_{xy} = H(x, y, 0) = \sum_i xh_i^x + yh_i^y,
\end{equation}
where $h_i^\eta = \frac{1}{2} \sum_i s_i^\eta s_{i+1}^\eta + s_i^\eta s_{i+2}^\eta + s_{i+1}^\eta s_{i+2}^\eta$. We can prove that the minimal energy of $h_i^\eta$ is $-1/8$ \cite{PhysRevB.100.125101} , thus the GSs energy $E_g \ge -(x+ y) L/8$, which can be reached by states 
in Eq. \ref{eq-wf}. With this model, we may construct a mixed Hamiltonian (see the general theorem for this decoupling in the Appendix),
\begin{equation}
H_{\text{x}} = \beta H(x',x',z') + (1-\beta)H(x'', y'', 0), 
\label{eq-mixed}
\end{equation}
where $z' > -x'/2$, $x'' > 0$, $y''> 0$ and $\beta \in [0, 1]$. We require that both $H(x',x',z')$ and $H(x'', y'', 0)$ have the same exact
dimer GSs of Eq. \ref{eq-wf}. Then, according to Eq. \ref{eq-b1}, we can find the exact dimer GSs when 
\begin{equation}
    \beta(2x'+z') +(1-\beta)(x''+y'') > 0,
\end{equation}
and 
\begin{equation}
	 \beta^2 (x'^2 +2x'z') + (1-\beta)^2 x''y'' + \beta(1-\beta) (x'' + y'')(x'+z') > 0,
\end{equation}
which can always be fulfilled for the given condition. So the decoupling of $H_{\text{x}}$ provides a general approach to construct exact 
dimer GSs from some simple (known) models, which can be used to understand the exact dimer states in some of the regions 
in the phase diagram of Fig. \ref{fig-fig2}. Nevertheless, not all regions in the phase diagram can be understood in this way. 
In Eq. \ref{eq-mixed}, one may replace the XXZ model by the anisotropic XYZ model and prove that this decoupling only allows 
solution when $z > -{1\over 2}\text{min}(\cos(\theta),\sin(\theta))$ for $\theta \in[0, \pi/2]$, 
$z > -2\cos(\theta)$ for $\theta \in(\pi/2, 
\pi-\arctan(2))$, and $z > -2\sin(\theta)$ for $\theta \in(3\pi/2+\arctan(2), 2\pi)$ (see the light red regions in 
Fig. \ref{fig-fig2}. The detailed analysis can be found in the Appendix. Beyond these three regions, the exact dimer phase 
can not be understood by the mechanism of Eq. \ref{eq-mixed}, which indicates of non-triviality for this phase.

\begin{figure}
\centering
\includegraphics[width=0.48\textwidth]{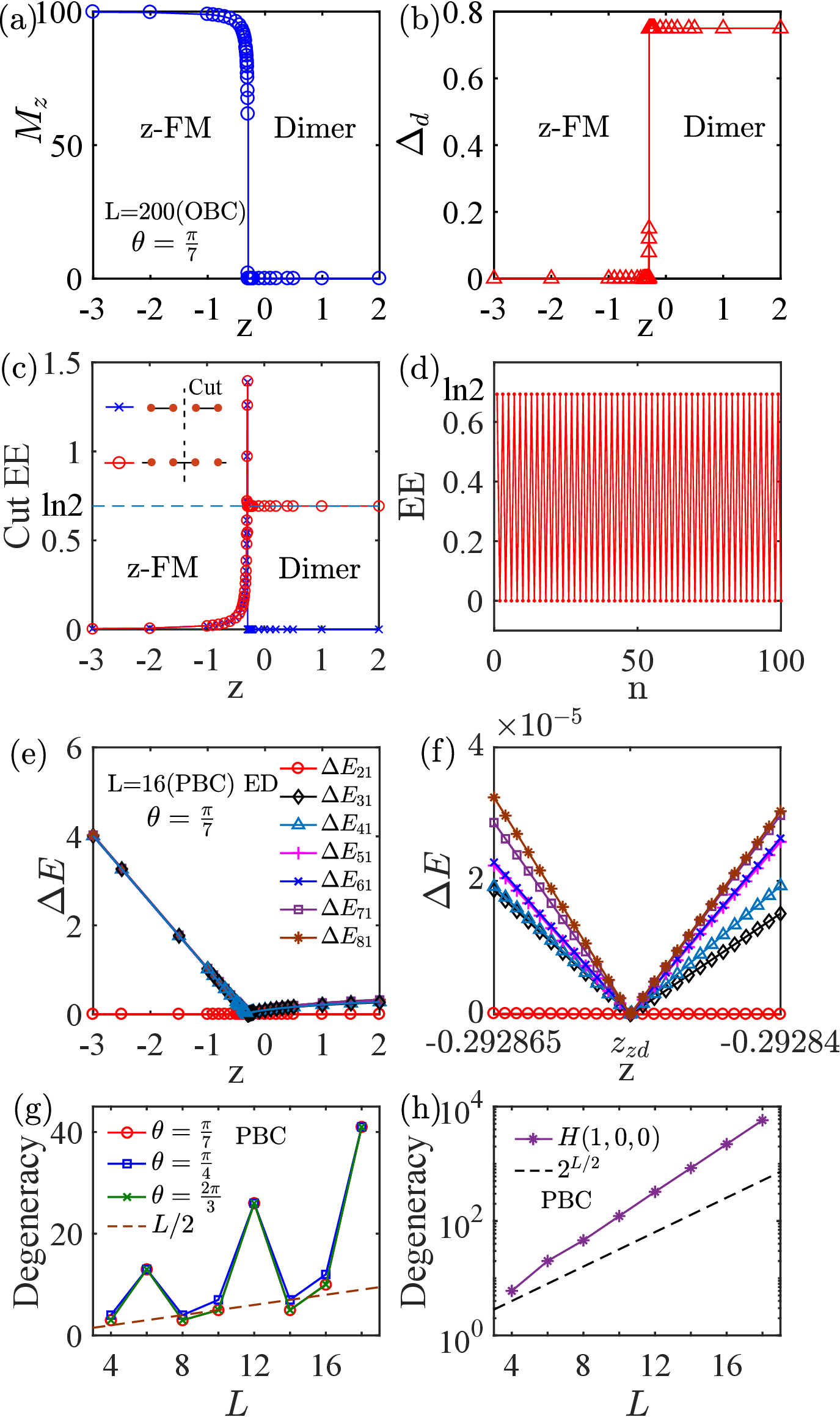}
	\caption{(a) Dimer and (b) magnetization orders at $\theta = \pi/7$ from density matrix renormalization group (DMRG) method with open boundary condition (OBC).  The numerical determined boundary is $z_{\text{ex}} = -0.29283$, while the exact boundary from Eq. \ref{eq-b1} is $z_c = -0.29285$. (c) The cut EE (see definition in the inset) as a function of $z$ at $\theta = \pi/7$.  At the phase boundary, the EE exhibits a sharp peak. (d) A typical result for oscillating of EE due to singlet dimer state. (e) Excitation gaps $\delta E_{n1}$ from $z$-FM to exact dimer phase. (f) The enlarged excitation gaps near the critical point. Data are obtained for $L= 16$ from exact diagonalization (ED) with PBC. (g) The degeneracy of the GSs at the phase boundary as a function of $L$ and $\theta$, which scales as 
$\mathcal{O}(L/2)$. (h) The degeneracy of the GSs of $H(1,0,0)$ with scaling of $\mathcal{O}(2^{L/2})$. }
\label{fig-fig4}
\end{figure}

\subsection{Order parameters and infinite-fold degeneracy} 

The boundary condition in Eq. \ref{eq-b1} automatically satisfies the permutation symmetry of $H$. This boundary is numerically verified with extraordinary high accuracy (see the dots in Fig. \ref{fig-fig2}). A typical transition from the exact dimer phase to the $z$-FM phase is presented in Fig. \ref{fig-fig4} (a) - (c), which is characterized by the
dimer order $\Delta_d$ \cite{furukawa2012ground,michaud2012antiferromagnetic}, magnetization  $M_\eta$ \cite{de2012entanglement} and entanglement entropy (EE). We define these two order parameters 
as
\begin{equation}
	\Delta_d = \langle {\bf s}_i \cdot {\bf s}_{i+1} - {\bf s}_{i+1}\cdot {\bf s}_{i+2}\rangle, \quad M_\eta = \sum_i \langle s_i^\eta\rangle.
	\label{eq-orders}
\end{equation}
Physically, the first order parameter reflects the translational symmetry breaking for dimerization; and the second one reflects the
time reversal symmetry breaking for the FM phases. To further characterize the entanglement feature, or quantumness, in these phases, 
we can calculate the EE of a finite block $A$ with size $n$, which is defined as \cite{holzhey1994geometric,junemann2017exploring,calabrese2009entanglement},
\begin{equation}
    S_A(n) = -\Tr \rho_A \ln \rho_A,
\end{equation}
where $\rho_A$ is the reduced density matrix by tracing out its complementary part. 
In the exact dimer phase, $\Delta_d = 3/4$, $M_z = 0$, and the central cut EE equals to 0 (at $n=L/2$) or $\ln 2$ 
(at $n=L/2+1$) due to formation of the singlet dimer state. In Fig. \ref{fig-fig4} (d), we show a typical result for oscillating 
of EE.  In the $z$-FM phase, $M_z - L/2 \propto 1/z^2$ (from second-order perturbation theory), $\Delta_d = 0$; and with 
the decreases of $z$, the cut EE tends to be zero when $z$ approaches the exact FM phase limit of $H(0,0,-1)$ (see 
section \ref{sec-fm} B). The boundary determined by these order parameters is precisely the same as that from  Eq. \ref{eq-b1}, 
with absolute difference less than $3\times 10^{-4}$. The similar accuracy has been found for all dots at the boundaries in 
Fig. \ref{fig-fig2}. In Fig. \ref{fig-fig4} (e) - (f), we show that at the phase boundary, the excitation gaps defined as 
$\delta E_{n1} = E_n - E_1$ for $n \geq 2$ collapse to zero, indicating of infinite-fold degeneracy when extending to infinite 
length. In Ref. \onlinecite{gerhardt1998metamagnetism}, Gerhardt {\it et al.} have proven that the infinite-fold degeneracy of the GSs at point $x=y$, $z=-x/2$ by considering the $n$-magnon states
\begin{equation}
S^{+}(p)^n |\text{FM}\rangle^z_{\text{exact}},
	\label{eq-nmagnon}
\end{equation}
which can be obtained by $n$-fold application of the raising operator $S^{+}(p)=\sum_l e^{ipl} s_l^+$.
Here, $|\text{FM}\rangle^z_{\text{exact}} = |\downarrow\rangle^{\bigotimes L}$ is FM state (see also the more general definition in Eq. \ref{eq-fm}).
One can see that the $n$-magnon states are eigenstates of the Hamiltonian
\begin{equation}
H(x,x,-\frac{x}{2}) S^{+}(p)^n |\text{FM}\rangle^z_{\text{exact}} 
= E_g^{\text{FM}} S^{+}(p)^n |\text{FM}\rangle^z_{\text{exact}},
\end{equation}
for $p=2\pi/3$ and $p=4\pi/3$, where the FM state energy is given by
\begin{equation}
	E_g^{\text{FM}} = -{3xL \over 16}. 
\end{equation}
This conclusion is achieved using 
\begin{equation}
\begin{aligned}
& [[H(x,x,-\frac{x}{2}), S^{+}(p)], S^{+}(p)]|\text{FM}\rangle^z_{\text{exact}} \\
= & -xe^{ip}[(1+2\cos(p))|2p,1\rangle + e^{ip}(\frac{1}{2} + \cos(2p))|2p,2\rangle ],
\end{aligned}
\end{equation}
where $|2p,j\rangle = \sum_l e^{2ipl}|l,l+j\rangle$ are the two-magnon states with two spin excitations at sites $l$ and $l+j$ (see Eq. \ref{eq-nmagnon}). The right-band side
disappears when $p = 2\pi/3$ and $p = 4\pi/3$.
At this point, the eigenvalues $E_g^\text{dimer}$ of the exact dimer states are degenerate with the energy $E_g^{\text{FM}}$ of the FM states, which  also implies
that the $n$-magnon states are GSs of $H(x,x,-x/2)$. Thus the GSs energies are degenerate with respect to total spin $S_z=0$, $1$, $2$, $\cdots$, $L/2$ sectors \cite{gerhardt1998metamagnetism}. Therefore, in the thermodynamic limit, the degeneracy of the GSs is at least of the order of $\mathcal{O}(L/2)$. In Fig. \ref{fig-fig4} (g), we show the degeneracy of the GSs at the phase boundary with PBC. We find that the degeneracy increases with some kind of oscillation from the finite size effect with the increasing of $L$, which scales as 
$\mathcal{O}(L)$. 

At the phase boundary, we also find three classical points $H(1,0,0)$, $H(0,1,0)$ and $H(0,0,1)$, with GSs degeneracy increases
exponentially with the increasing of system size $L$. 
Here, $H(0,0,1)$ is relevant to the boundary defined in Eq. \ref{eq-b1} in the limit of $x = -y$ 
and $z \rightarrow \infty$. Let us consider $H(x,0,0) = xH(1,0,0)$ for $x > 0$, and \cite{PhysRevB.100.125101}
\begin{equation}
    H(1, 0, 0) =\frac{1}{2} \sum_i^L \sigma_i \sigma_{i+1} - \frac{L}{8}, \quad \sigma_i = \{-1, 0, 1\},
	\label{eq-H100}
\end{equation}
where $\sigma_i = s_i^x + s_{i+1}^x$. This new operator takes three different values; however, the minimal value $-1$ from the product of the operators can not be reached due to the restriction 
$|\sigma_i - \sigma_{i+1}| = |s_i^x - s_{i+2}^x| = \{0, 1\}$. Thus $\sigma_i\sigma_{i+1} \ge 0$ and the GSs energy is $E_g = 
-L/8$. Let us consider a special case, that is, $\sigma_{2i} = 0$,
and $\sigma_{2i+1} = \{1, 0\}$ or $\{-1,0\}$. All these states have the same GSs energy $E_g = -L/8$. 
This means that the degeneracy of the GSs is at least of the order of 
$\mathcal{O}(2^{L/2})$, which is infinite-fold degenerate in infinite length (see verification in 
Fig. \ref{fig-fig4} (h)). From this boundary, the system may undergo two different spontaneous symmetry breakings. 
When it breaks to the exact dimer phase, the system breaks the translational symmetry with $\Delta_d \ne 0$; while to the FM phases, 
it breaks the $\mathbb{Z}_2$ symmetry with $M_\eta \ne 0$ and $\Delta_d =0$. Since we have three different 
$\mathbb{Z}_2$ operators for symmetry breaking, we have three different FM phases. 

\section{Ferromagnetic phases}
\label{sec-fm}

We find three different FM phases polarize along the three orthogonal directions $x$, $y$ and $z$. From the point of view of
symmetry breaking, these FM phases correspond to the spontaneous $\mathbb{Z}_2$ symmetry breaking along the three axes. 
The transitions between them are phase transitions and 
the boundaries are gapless and critical. The three boundaries for the FM phases are $z=x$ for $\theta\in(\pi/2,5\pi/4)$, $z=y$ for $\theta\in(5\pi/4,2\pi)$, and $z > x=y = -1/\sqrt{2}$.  Across these boundaries, the polarization of magnetizations will change 
direction. In the following, we use several complementary approaches to characterize these phase transitions. 

\begin{figure}
\centering
\includegraphics[width=0.50\textwidth]{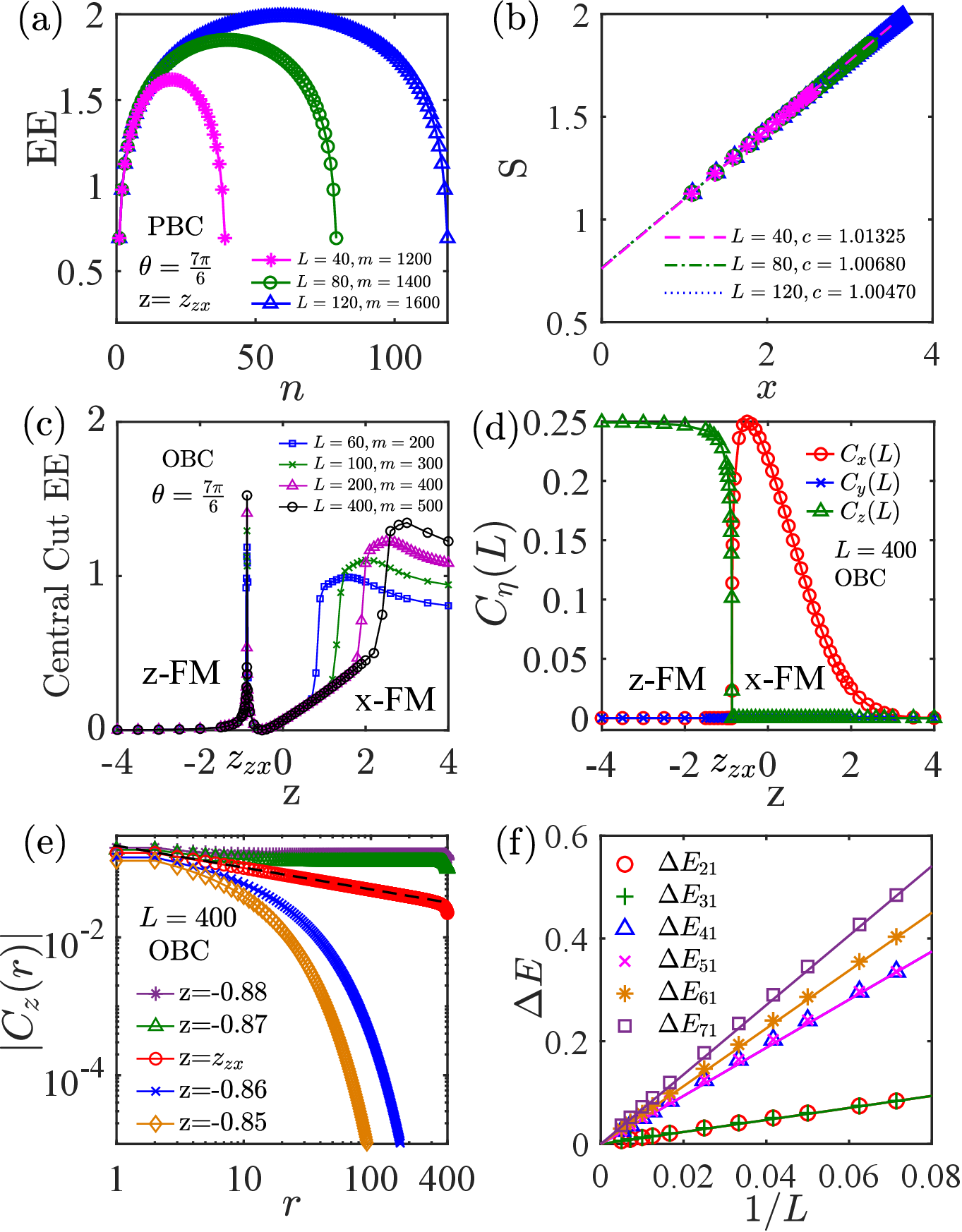}
\caption{
 (a) and (b) show EE and central charge $c$ at the boundary between $z$-FM and $x$-FM phase at $\theta = 7\pi/6$ with  $z_{\text{zx}} = -0.86602$, by DMRG method with PBC. The dashed lines in (b) are fitted by Eq. \ref{eq-Sn} with $x=\ln(\frac{L}{\pi}\sin\frac{\pi n}{L})$, yielding $c = 1$. (c) Central cut EE as a function of $z$ at $\theta = 7\pi/6$ for different $L$ and bond dimension $m$. (d) Spin-spin correlation functions $C_\eta(L)$ ($\eta=x,y,z$) as a function of $z$ at $\theta = 7\pi/6$ for $L=400$.  (e) Spin-spin correlation function $C_z(r)$ at $\theta = 7\pi/6$. At the phase boundary, $C_z(r) \propto |r|^{-0.32}$, by DMRG method with OBC. (f) Scaling of excitation gaps $\delta E_{n1} \propto 1/L$ for all $n$ at the boundary ($\theta = 7\pi/6$ with $z_{\text{zx}} = -0.86602$) as a function of chain length, indicating of gapless and criticality.}
\label{fig-fig5}
\end{figure}

\subsection{Properties of FM phases}
The phase boundaries of the three FM phases can be obtained by performing the dual transformation
\begin{equation}
\mathcal{R}_\eta = \prod_i \exp(i\frac{\pi}{2} s_i^\eta), \quad \eta=x,y,z.
\end{equation}
For example, by performing $\mathcal{R}_y$, $H(x,y,z)$ is transferred to $H(z,y,x)$. This transformation means that the total Hamiltonian is invariant when $z=x$. Therefore, the boundaries are self-dual lines, which are gapless and critical. 
In order to verify these boundaries, we consider the EE in a finite chain with PBC as \cite{holzhey1994geometric,junemann2017exploring,calabrese2009entanglement},
\begin{equation}
	S_A(n) = {c \over 3} \ln({L \over \pi} \sin {\pi n \over L}) +s_0,
	\label{eq-Sn}
\end{equation}
where $c$ refers to central charge and $s_0$ is a non-universal constant. The results are presented in Fig. \ref{fig-fig5} (a) - (b). 
We find that the central charge $c = 1$ at the phase boundary, which is a typical feature of free fermions. In Fig. \ref{fig-fig5} (c), we show the central cut EE defined
as $S(L/2)$ as a function of $z$ at $\theta = 7\pi/6$ for different $L$ and bond dimension $m$.  At the phase boundary, we find 
that the EE exhibits a sharp peak, and its value increases with the increasing of lattice site $L$, reflecting gapless and 
criticality. In the $z$-FM phase, with the decreasing of $z$, it will approach the exact FM phase limit $H(0,0,-1)$, so the central cut EE tends to zero. 
However, in the $x$-FM phase, as $z$ increases, the central cut EE first decreases (at the exact FM state point $z=\sin(7\pi/6)$, it equals to zero) and then increases (close to the infinite-fold degeneracy point $H(0,0,1)$); see details in section \ref{sec-exactdimer}-B.

This phase transition may also be characterized by their long-range spin-spin correlation functions
\begin{equation}
C_\eta(r) = \langle s_1^\eta s_{r}^\eta\rangle, \quad \eta=x, y, z.
\end{equation}
In Fig. \ref{fig-fig5} (d), we show the $C_\eta(L)$ as a function of $z$ at $\theta = 7\pi/6$ for $L=400$. As expected, in the $z$-FM phase, $C_{x,y}(L)=0$ and $C_z(L)\neq 0$, while in the $x$-FM phase, $C_{y,z} = 0$ and $C_x(L)\neq 0$. In Fig. \ref{fig-fig5}  (e), we study the correlation function $C_z(r)$ near the phase boundary. In the fully gapped $z$-FM phase with long-range order, this correlation function approaches a constant in the large separation limit. At the boundary, $C_z(r) \propto |r|^{-\gamma}$, which is a typical feature of critical phase. In the $x$-FM phase with spin polarization along $x$-direction, the correlation function $C_z(r)$ decays exponentially to zero; on the contrary, $\lim_{|r| \rightarrow \infty} C_x(r)$ approaches a constant. 

We also study the excitation gaps, which is defined as the energy difference between the excited states and the ground state as
\begin{equation}
\Delta E_{n1} = E_n-E_1 = \Delta E_{n1}(\infty) + \frac{A_n}{L}, \quad n=2,3,\cdots.
\end{equation}
At the phase boundaries, we find $\Delta E_{n1}(\infty) = 0$, which also 
means that the boundaries are gapless and critical (see Fig. \ref{fig-fig5} (f)). These features are consistent with 
the finite central charge ($c = 1$) observed from central cut EE. 

\subsection{Exact FM states}

There exist some special lines in the FM phases to support exact FM states as \cite{exactfm}
\begin{equation}
	|\text{FM}\rangle_\text{exact}^\eta = |\eta\rangle^{\otimes L}, 
        \label{eq-fm}
\end{equation}
where $|\eta\rangle$ is the eigenvector of $s^{\eta}$. As shown in Ref.  \onlinecite{gerhardt1998metamagnetism}, when $y=x > 0$ and $z < -x/2$, the ground state is an exact FM 
state spontaneously polarized along $z$-direction (thus breaks $\mathbb{Z}_2$ symmetry along $z$ axis). In our model, we also find another exact $z$-FM phase when $z < x=y = -1/\sqrt{2}$. This state can be mapped to the exact FM state along the other two directions by dual rotation $\mathcal{R}_{x,y}$, which induces permutation among the three directions. We find that the other two exact FM states at 
$z = y$ for $\theta\in(\pi-\arctan(2),5\pi/4)$ and $z = x$ for $\theta\in(5\pi/4,3\pi/2+\arctan(2))$. These special cases 
are presented in Fig. \ref{fig-fig1} with dashed lines, in which the arrows mark the evolution of these dual mapping starting 
from $z \rightarrow -\infty$. One should be noticed that when $z \rightarrow -\infty$, it equals to $-H(0,0,1)$, which can be mapped to $-H(1,0,0)$ 
and $-H(0,1,0)$ by dual rotations. The GSs of these points should be two-fold degenerate with exact FM states in Eq. \ref{eq-fm}. This exact two-fold degeneracy can also be proven 
exactly by considering $-H(0, 0, 1)$ using the method in Eq. \ref{eq-H100}.  In these exact FM states, the corresponding ground state energy can be written as 
\begin{equation}
E_g^{\text{FM}} = -{3L \over 8}| \text{min}\{x, y, z\}|.
\end{equation}
Notice that the GSs of $-H(1,1,1)$ is infinite-fold degenerate, while in $H(1,1,1)$ it is exactly dimerized. This may provide 
an interesting example for the relation between the wave functions of the GSs and the highest energy states.

\section{Experimental relevant and measurements}
\label{sec-app}

Let us finally discuss the relevance of this research to experiments in one dimensional magnets and their possible 
experimental signatures. The results in the previous sections are demonstrated at the exact MG point for the sake of 
exact solvability; however, these physics can be survived even when slightly away from this point, which can happen in 
real materials. These states are still characterized by the order parameter $\Delta_d \ne 0$ with a finite energy gap;  
however, it is no longer the exact dimer phase discussed before with wave function given in Eq. \ref{eq-wf}. 
These physics needs to be explored using numerical methods. 
In the spin-Peierls compounds, such as CuGeO$_3$ \cite{o2017vibronic}, TiOCl \cite{rotundu2018enhancement} and (TMTTF)$_2$PF$_6$ \cite{pouget2017inelastic}, the strong anisotropy in lattice constants 
(for example, in CuGeO$_3$ the lattice constants are: 
$a=8.4749$ $\AA$, $b=2.9431$ $\AA$ and $c=4.8023$ $\AA$ \cite{meibohm2018comparison}) is necessary to isolate a single Cu$^{2+}$ chain (or other spin-${1\over 2}$ ions) 
out from the three-dimensional bulk. For this reason, spatial anisotropy is inevitable and in order to describe real materials more accurately, anisotropy in the effective spin model is needed. In experiments, it was found that when the temperature is lower than the spin-Peierls transition 
temperature $T_{\text{sp}}$, the magnetic susceptibility in all directions will quickly drop to almost zero. Anisotropy in magnetic susceptibility will become significant in the FM phase when the Zeeman field exceeds a critical value or $T > T_\text{sp}$. In experiments, these observations are explained by an isotropic $J_1$-$J_2$ model, which may support the dimer
phase when $\alpha=J_2/J_1 > 0.2411$ \cite{bursill1995numerical}. This isotropic model was also shown to relevant to other anisotropic one dimensional magnets such as CuCrO$_4$ with $\alpha = 0.43$ \cite{law2011quasi}, BaV$_3$O$_8$ with $\alpha \approx 0.5$ \cite{chakrabarty2013bav}, 
Cu$_3$(MoO$_4$)(OH)$_4$ with $\alpha = 0.45$ \cite{lebernegg2017frustrated}, Cu$_6$Ge$_6$O$_{18}\cdot 6$H$_2$O with $\alpha = 0.27$  \cite{hase2003magnetic}, 
Cu$_6$Ge$_6$O$_{18} \cdot$H$_2$O with $\alpha = 0.29$ \cite{hase2003magnetic} and Li$_{1.16}$Cu$_{1.84}$O$_{2.01}$ with $\alpha = 0.29$ \cite{masuda2004competition}. In some of the experiments, anisotropy has been reported. For instance, in CuGeO$_3$ in Refs. \onlinecite{miljak2005anisotropic, wang2013single, Kremer1995Anisotropic}, the measured spin susceptibilities 
along the three crystal axes directions are different, differing by about 10 - 20\%, and the parameters are determined to be $\alpha=0.71$, $J_{xx} = 48.2$ K, $J_{yy}=47.2$ K and $J_{zz} = 49.7$ K. 
In some materials, these parameters may even be negative valued. These results motivate us to think more seriously about the dimer phase in these compounds. 

\begin{figure}
    \centering
    \includegraphics[width=0.50\textwidth]{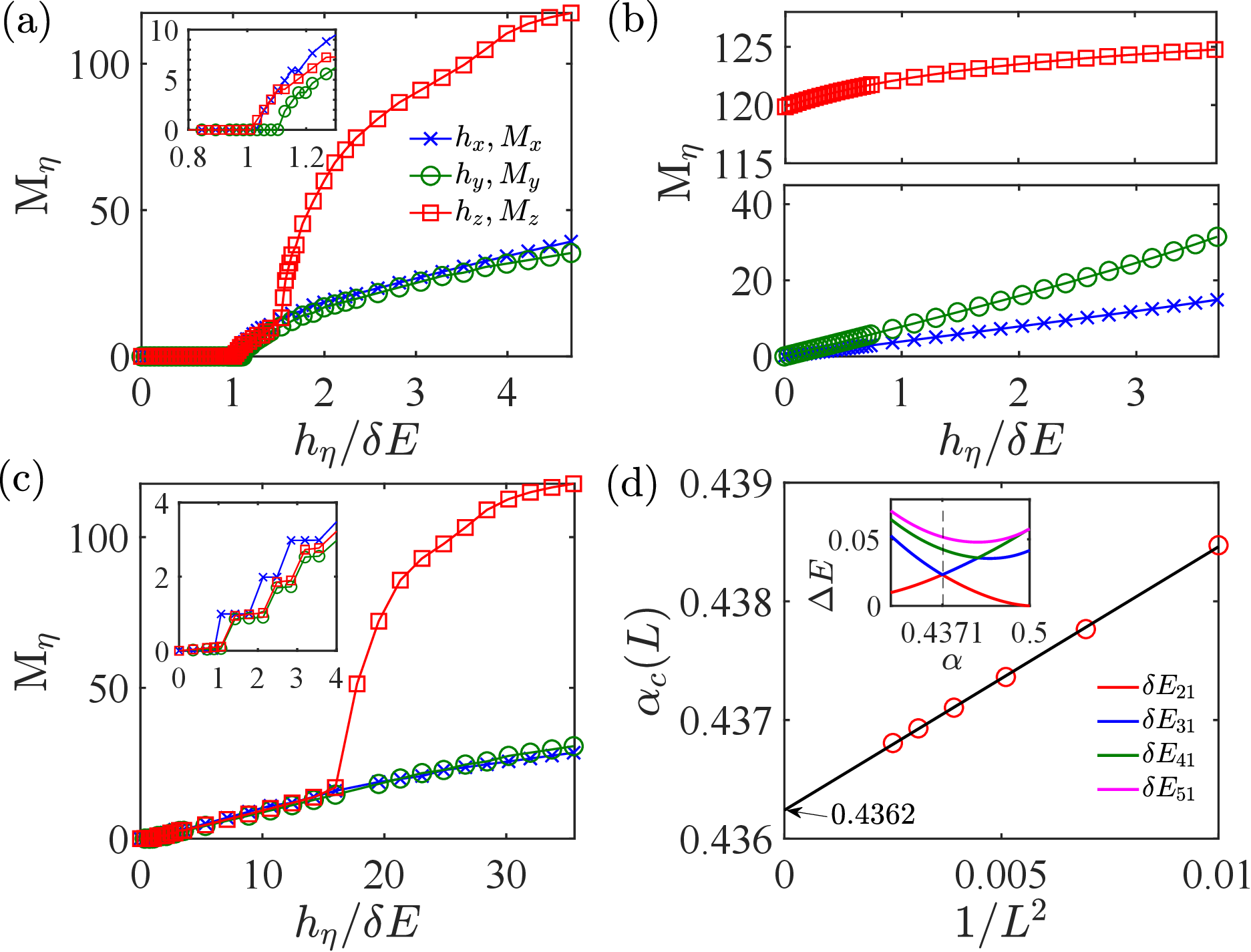}
	\caption{Magnetizaton $M_\eta$ at $\theta=\frac{\pi}{7}$. (a) Exact dimer phase with $\alpha={1\over 2}$, $z=-0.2$. The three critical Zeeman fields are 
    $h_x^c = 0.044$, $h_y^c = 0.047$ and $h_z^c=0.043$, and excitation gap $\delta E=0.0425$. 
    (b) $z$-FM phase with $\alpha=\frac{1}{2}$, $z=-0.4$, $\delta E=0.054$.
    (c) Dimer phase with $\alpha=0.45$, $z=-0.2$, $h_x^c = 0.005$, $h_y^c = 0.006$ and $h_z^c=0.006$, $\delta E=0.0056$.
	 These results are obtained with $L=256$ based on DMRG method. (d) Critical boundary for dimer phase at $\theta=\frac{\pi}{7}$ and $z=-0.2$. The critical point $\alpha_c = 0.4362$ is obtained by extrapolating to infinity length (see Eq. \ref{eq-DeltaL}). Inset shows the boundary determined by level crossing between the first and second excited states \cite{somma2001phase}.}
    \label{fig-fig6}
\end{figure}

We model the experimental measurements by adding a magnetic field along $\eta$-direction,
\begin{equation}
    H' = h\sum_i^L s_i^\eta, \quad \eta  =x, y, z. 
\end{equation}
Since there is an energy gap $\delta E = E_3-E_1$ in the exact dimer phase (note that $E_1=E_2$ for $\mathbb{Z}_2$ symmetry), 
the external magnetic field will not immediately destroy the exact dimer phase. The magnetization $M_\eta$ (see Eq. \ref{eq-orders}) 
for the exact dimer phase along different directions are presented in Fig. \ref{fig-fig6} (a). We find that the breakdown of the exact dimer phase takes place roughly at 
\begin{equation}
h^c \simeq \delta E = E_3 - E_1.
\end{equation}
When $h < h^c$, the magnetization $M_\eta=0$ along different directions, thus $\chi_\alpha = \partial M_\alpha /\partial h  = 0$. The anisotropy effect will be 
important in the region when $h > h^c$ or $T> T_{\text{sp}}$, which gives different susceptibilities for the magnetic field along different 
directions. This result is consistent with the experimental observations \cite{ryu2016surface, zayed20174, yamaguchi2013various,
sahling2015experimental,lebernegg2017frustrated}. This anisotropy effect has been reported even in the first spin-Peierls compound $\text{CuGeO}_3$ 
\cite{hase1993observation}. In Fig. \ref{fig-fig6} (b), we show the magnetization $M_\eta$ in the $z$-FM phase and find that even a small $h$ can lead to a non-zero $M_\eta$. These features can be used to distinguish these fully gapped dimer and FM phases.
In Fig. \ref{fig-fig6} (c), we plot the magnetization away from the MG point. The similar features can also be found in the dimer phase, and the phase transition can still take place at $h^c \simeq \delta E$. 
In experiments, the value of $\alpha$ depends strongly on the lattice constants, thus maybe tuned by temperature or external stress \cite{martin2011larmor}. We plot the boundary for the dimer phase at $\theta=\frac{\pi}{7}$ and $z=-0.2$ in Fig. \ref{fig-fig6} (d), yielding 
$\alpha_c = 0.4362$ by extrapolation to infinite length using \cite{somma2001phase, riera1995magnetic}
\begin{equation}
\alpha_c(L) - \alpha_c \propto L^{-2}.
	\label{eq-DeltaL}
\end{equation}
We have also confirmed this boundary from the dimer order $\Delta_\text{d}$ and the long-range correlation function $C_{\eta}(r)$. This critical value is significantly larger than $0.2411$ in the isotropic $J_1$-$J_2$ model from the anisotropy effect. From the general 
theorem demonstrated in this work, we see that the dimer phase in the isotropic model will be survived even in the presence of 
anisotropy, thus it may find applications in the above materials.

\section{Conclusion}
\label{sec-conc}

This work is motivated by the generalization of the isotropic MG model to the anisotropic region, which may have applications in
realistic one dimensional magnets. We demonstrate that the exact dimer phase can be found in a wide range of parameters 
in a generalized MG model (at point $\alpha = {1 \over 2}$) with anisotropic XYZ interaction. Due to the presence of the exact dimer
phase, the phase boundaries can be obtained analytically using simple models, which are verified with high accuracy using 
some numerical methods. We find that this model support one exact dimer phase and three FM phases, which polarize in different 
directions. The boundaries between the exact dimer phase and FM phases are infinite-fold degenerate, 
while between the FM phases are gapless, critical with central charge $c = 1$ for free fermions. These 
results may be relevant to a large number of one dimensional magnets. Possible signatures are presented to discriminate 
them in experiments. These results may advance our understanding of dimer phases in solid materials, and it may even have application in the searching of SPT phases 
\cite{chen2011classification,chen2013symmetry,chen2012symmetry,chen2011complete} from the general theorem proven in this manuscript.

\setcounter{equation}{0}
\renewcommand\theequation{A\arabic{equation}}

\section*{APPENDIX}

The major results in this work are established based on the following general theorem. 

{\it Theorem}. When the Hamiltonians $H_i$ have the same GS wave functions, then these GSs will also be the GSs of the 
Hamiltonian $H = \sum_i H_i$, in which these Hamiltonians $H_i$ are not necessarily commutative to each other. 

We first prove this theorem using two Hamiltonians $H_1$ and $H_2$, with corresponding GS wave functions as 
$|g_1\rangle$ and $|g_2\rangle$. Let us define $H = H_1 + H_2$, with GS as $|g\rangle$. Then
\begin{eqnarray}
	\langle g|H| g\rangle && =\left\langle g\left|H_{1}+H_{2}\right| g\right\rangle=\left\langle g\left|H_{1}\right| g\right\rangle+\left\langle g\left|H_{2}\right| g\right\rangle \nonumber \\ 
						  &&  \geq \left\langle g_{1}\left|H_{1}\right| g_{1}\right\rangle+\left\langle g_{2}\left|H_{2}\right| g_{2}\right\rangle.
\label{eq-A1}
\end{eqnarray}
In particular when $H_1$ and $H_2$ have the same GSs that $|g^\prime\rangle = |g_1\rangle = |g_2\rangle$, then Eq. \ref{eq-A1} can be further written as
\begin{equation}
\begin{aligned}
\langle g|H| g\rangle & \geq \left\langle g^\prime\left|H_{1}\right| g^\prime\right\rangle+\left\langle g^\prime\left|H_{2}\right| g^\prime\right\rangle \\
& = \left\langle g^\prime\left|H_{1}+H_{2}\right| g^\prime\right\rangle = \left\langle g^\prime\left|H \right| g^\prime\right\rangle.
\end{aligned}
\end{equation}
This means that $|g^\prime\rangle$ is the ground state of $H$. This conclusion can be generalized to an arbitrary number of Hamiltonians $H_i$ with $i=1$, $\cdots$, $ n$ 
for $n > 2$, in which $H = \sum_i H_i$ will have the same GSs as $H_i$. In the main text of Eq. \ref{eq-mixed}, the 
GSs of $H(x^\prime,x^\prime,z^\prime)$ and $H(x^{\prime\prime},y^{\prime\prime},0)$ are exact dimerized with wave functions 
defined in Eq. \ref{eq-wf}, so the GSs of $H_{\text{x}}$ should also be the same dimerized states, following this general 
theorem.

It should be emphasized that in the above theorem, the different Hamiltonians $H_i$ are not necessarily to be commutative 
to each other. In the main text, we have used two different Hamiltonians $H_1= H(x^\prime,x^\prime,z^\prime)$ and 
$H_2 = H(x^{\prime\prime},y^{\prime\prime},0)$, which have the same GSs; however, $[H_1, H_2] \ne 0$. For this reason, this theorem should be different from the concept of the complete set of commuting observables (CSCO) used in textbooks, which state that when two 
operators $H_1$ and $H_2$ are commutate, then they will have the common eigenvectors; however, these two Hamiltonians may have different GSs. For instance, if we define a folded Hamiltonian as $H_2 = (H_1 - \epsilon)^2$. Apparently, $[H_2, H_1] = 0$. However, the GS of $H_2$ may correspond to the excited state of $H_1$ 
with energy closes to $\epsilon$. Another example, which is much simpler, is $H_2 = -H_1$, in which the GS of $H_1$ is the
highest energy state of $H_2$; and vise versa. 

This theorem can be used to explain a part of the exact dimer phase in the phase diagram (see the light red regions in Fig. \ref{fig-fig1}), but not all. We can define 
$x = \cos(\theta)$ and $y = \sin(\theta)$ (see Fig. \ref{fig-fig2}), then we have the following three cases. These cases have 
been verified by numerical simulation. We find that including much more decouplings will not change these conclusions; 
in  other words, the deep red regions in Fig. \ref{fig-fig2} can not be explained by this decoupling. 

(1) For $\theta\in[0,\pi/2]$; if $z>0$, we have
decoupling as 
\begin{equation}
H(x,y,z) = H(\frac{x}{2},\frac{y}{2},0) + H(\frac{x}{2},0,\frac{z}{2}) + H(0,\frac{y}{2},\frac{z}{2}). 
\end{equation}
However, if $z \leq 0$, we have
\begin{equation}
H(x,y,z) = H(-2z,-2z,z) + H(x+2z,y+2z,0),
\end{equation}
where $H(-2z,-2z,z)$ locates at the phase boundary of XXZ model. Therefore, if the GSs of $H(x,y,z)$ are exact dimer states, $H(x+2z,y+2z,0)$ should satisfy $x+2z>0$ and 
$y+2z>0$, which yields 
\begin{equation}
z>-\frac{1}{2}\min(\cos(\theta),\sin(\theta)).
\end{equation}

(2) For $\theta\in(\pi/2,\pi]$, we can make a unitary transformation $\mathcal{R}_y$, which transfers $H(x,y,z)$ to $H(z,y,x)$, then
\begin{equation}
H(z,y,x) = H(-2x,-2x,x) + H(z+2x,y+2x,0).
\end{equation}
In order to ensure that the GSs of $H(x,y,z)$ are exact dimer states, the condition $z+2x>0$ and $y+2x>0$ need to be satisfied, yielding
\begin{equation}
\theta\in(\pi/2,\pi-\arctan(2)), \quad z>-2\cos(\theta).
\end{equation}

(3) For $\theta\in[3\pi/2,2\pi]$, the exact dimer states can be obtained in a similar method by a unitary transformation $\mathcal{R}_x$, the exact dimer phase requires 
\begin{equation}
\theta\in(3\pi/2+\arctan(2),2\pi), \quad z>-2\sin(\theta).
\end{equation}
The above three conditions have been used at the end of section \ref{sec-exactdimer}-A, indicating that the deep red regions 
in the phase diagram can not be understood based on this approach.

This general theorem may have much broader applications because the validity of this theorem can be applied to diverse physical 
models, including the single-particle models as well as the many-body models. It only requires that the Hamiltonians $H_i$ have 
the same GSs, so it may also have potential application in the construction of SPT phases. 

\begin{acknowledgments}
This work is supported by the National Key Research and Development Program in China (Grants No. 2017YFA0304504 and No. 2017YFA0304103) and the  National Natural Science Foundation of China (NSFC) with No. 11774328.

\end{acknowledgments}

\bibliography{ref}

\begin{thebibliography}{77}
\bibitem [{exa()}]{exactfm}%
  \BibitemOpen
  \href@noop {} {}\bibinfo {note} {By assuming the exact state to be
  $((\cos(\theta/2)e^{i\phi/2}, \sin(\theta/2)e^{i\phi/2})^T)^{\otimes L}$ and
  minimize the corresponding ground state energy with respect to parameters
  $\theta$ and $\phi$, we find that for the exact FM phase, the spin can only
  polarize along $x$, $y$ and $z$ directions.}\BibitemShut {Stop}%
\end{thebibliography}

\end{document}